# Discharge properties of a magnetized cylindrical capacitively coupled plasma discharge

Akanshu Khandelwal[a,b], Dhyey Raval[c], Narayan Sharma[b], Yashashri Patil[c], Sarveshwar Sharma[c,d], Shantanu Karkari[c,d] and Nishant Sirse[a,b]

[a] *Institute of Science and Research, IPS Academy, Indore-452012, India*
[b] *Centre for Scientific and Applied Research, IPS Academy, Indore-452012, India*
[c] *Institute for Plasma Research, Bhat, Gandhinagar, Gujarat, 382428 India*
[d] *Homi Bhabha National Institute, Training School Complex, Anushaktinagar, Mumbai-400094, India*

Corresponding author: N. Sirse

*E-mail address:* nishantsirse@ipsacademy.org

## Abstract

This work investigates the discharge properties of a cylindrical magnetized capacitive coupled plasma discharge produced between a pair of coaxial cylinders. For the purpose of diagnosing plasma properties and electron energy distribution function (EEDF), an in-house electronic circuit and an RF-compensated Langmuir probe are devised and constructed. The second harmonic technique (SHT) has been applied to obtain the direct measurement of EEDF and evaluated the effect of RF electron magnetization on the bulk plasma heating within the discharge. The effect of external magnetic field on the overall rise in plasma properties has been verified by use of particle and energy balance equations derived for argon discharge.

## 1. Introduction

In the plasma processing applications, radio frequency discharges are preferred over direct current (DC) discharges since they produce higher plasma density of the order of ~ $10^{15}$-$10^{17}$ m$^{-3}$ comparatively at a lower RF power [1,2]. RF discharges like capacitively coupled plasma (CCP) or Inductively coupled plasma (ICP) are preferred owning to its simple configuration, radial plasma uniformity and ion directivity towards the substrate to be processed as required in the case of deep reactive ion etching [3]. These sources operate in industrial scientific and medical (ISM) frequency band which lies between the ion plasma frequency ($\omega_{pi}$) and electron plasma frequency ($\omega_{pe}$) i.e. $\omega_{pi} \leq \omega_{rf} \ll \omega_{pe}$, where $\omega_{rf}$ is the applied RF frequency. In such scenario, the electrons are mostly responsible for absorbing power from the applied RF field, whereas heavier ions respond to time-averaged field present in the sheath region leads to the energy gain during their transit towards the electrodes.

Parallel plate configuration is the most widely used electrode design in CCP discharges. One of the major requirements in such discharges is to enhance and control plasma density independently compared to ion energy. In a single frequency plasma excitation, both the plasma density and ion energy scales with RF power [4–9]. Multiple frequency excitation [10–20] and tailored waveform [21–26] provides a level of tuning over plasma density and ion energy independently. On the other hand, obtaining high plasma density discharges always remains a challenge. Very high frequency plasma excitations were proposed to enhance the plasma density in the discharge [27–35].

External magnetic field is an another effective way to increase the efficiency of RF CCP discharges [36–47]. It has been observed that when a small magnetic field perpendicular to the RF electric field is applied, it enhances plasma density by confining electrons due to magnetic force [42–45]. You *et al.* after various simulation and experimental studies, has established that the effect of magnetic field is comparable as increase in the background gas pressure and produces non-local to local conversion of the plasma kinetics [48–50]. Several simulation and experimental studies have been performed to explore the effect of magnetic field on plasma heating [38,48,49,51]. Turner *et al.* in a study, demonstrated that the magnetic field reduces the phase difference between RF current and voltage thereby making the plasma highly resistive resulting an improvement in the power transfer efficiency and enhancing the plasma density [52]. Patil et al. [43] demonstrated that electron bounce resonance phenomenon could lead to a higher plasma density in VHF CCP discharge even at lower magnetic field strengths. Similar results were obtained by Zhang et al.[44] in the frequency range of 13.56-60 MHz where higher plasma densities were measured in both simulation and experiments.

Although the transverse magnetic field strength increases the plasma density in parallel plate CCP, the non-uniformity arises in the plasma density profile due to E×B drift [53]. Shin *et al.* performed reactive ion etching in parallel plate CCP discharge and observed application of static field generates charging current of opposite polarities in the two half of the wafer due to E×B drift [54]. Besides non uniformity in density, E×B drift also induces non uniformities in the electron temperature. S. Binwal *et al.* studied the effect of transverse magnetic field on the distribution of electron temperature in parallel plate CCP discharge [55]. It was observed that in the presence of magnetic field, the collision and hence the Ohmic heating near to the sheaths increases. As a result, the average electron temperature of the plasma increases.

Another novel approach to reduce lateral inhomogeneities is the use of cylindrical electrodes combined with an axisymmetric magnetic field. In such devices, by powering the cylindrical electrodes, the electric field produces in the radial direction and applied magnetic field in the axial direction generate magnetic force in the azimuthal direction. As a result of E×B force the electrons gets confined thereby increasing the ionization as well as improving the radial homogeneity. Experimental and theoretical works were reported related to the cylindrical CCP source with the axisymmetric magnetic field. Yeom *et al.* produced the cylindrical post-magnetron plasma and explained the dependency of self-bias on the applied magnetic field [56]. Lieberman *et al.* solved the transport model for cylindrical and coaxial electrode systems and showed good agreement with experimental results [57]. Joshi *et al.* used an arrangement of three cylinders along the axis together with an axial magnetic field to produce the plasma and they have observed that cylindrical electrodes in presence of axial magnetic field compliment the plasma uniformity due to closed azimuthal (E×B) drift [58]. Swati *et al.* produced the plasma inside a cylindrical electrode with two annular rings at both the ends of the electrode in the presence of an axial magnetic field. They have observed that an axially flowing constant net positive flux at different radial positions [46,59]. In a recent study performed in electronegative plasma[60], it was demonstrated that such discharges are efficient in negative ions production for a specific range of magnetic field strength. Although the radial uniformity and density profile improves, such plasma sources are geometrically asymmetric due to large cylindrical electrode.

In order to remove the geometrical asymmetry, understanding electron heating though RF sheaths, plasma asymmetry and validation of the simulation models, a nearly geometrically symmetric CCP source with axisymmetric magnetic field is envisaged. A novel, nearly symmetric, capacitive-coupled radiofrequency (RF) plasma device with axisymmetric magnetic field is developed in which plasma is produced in between two coaxial cylinders. To the best of our knowledge, no systematic studies were reported related to the properties of the capacitively coupled annular plasma with axisymmetric magnetic field. An advantage of such discharge is that the annular gap between the two coaxial cylinders is fixed and hence is equivalent to parallel plate CCP discharge. At the same time, the presence of axisymmetric magnetic field together with the radial electric field generates closed E×B drift in azimuthal direction. Due to this reason, the electrons get trapped in this bounded annular region. As a result, the plasma density could be enhanced in such devices, which is useful for various applications including ion source.

The paper is structured as follows. In section 2, the design, development and fabrication of plasma source, electromagnets, RF-compensated Langmuir probe and associated electronic circuit for the EEDF measurement are described in detail. In section 3, under results and discussion, we have presented the outcome of the experiment together with the detailed discussion of the result. Finally, the paper is concluded with the summarization and essential outcomes of the studies.

## 2. Plasma Source and Diagnostics

### 2.1 Plasma Source

Fig. 1 shows the schematic of the coaxial annular cylindrical CCP discharge set-up consisting of a plasma chamber, electrode assembly, electromagnet system, pumping unit and RF power supply. The experimental set up consists of a main vacuum chamber of 100 cm length and 27.6 cm diameter. The vacuum chamber is pumped down to a base pressure of $5\times10^{-6}$ mbar by using a diffusion pump backed by rotary pump. The pressure is measured by a full range pressure gauge attached to one of the radial-ports of the chamber. The electrode consists of two co-axial stainless-steel cylinders kept inside this vacuum chamber as shown in Fig.1. The inner cylinder has an outer diameter of 14.2 cm and outer cylinder has an inner diameter of 24.2 cm, providing a discharge gap of 5 cm. Length of both the cylinders is 15 cm. Inner cylinder is powered by using 13.56 MHz frequency RF power supply (Coaxial Power systems-AG1213W) via matching unit and outer cylinder is grounded. Both the cylinders are isolated from each other as well as from the main chamber by using high frequency insulator blocks. The plasma is produced in the annular region bounded by these two co-axial cylindrical electrodes. For axial confinement, floating confinement grids are used at both the ends of the cylindrical electrodes as shown in Fig.1(a). Fig. 1 (b) shows argon plasma in the annular region between co-axial cylinders confined between axial grids.

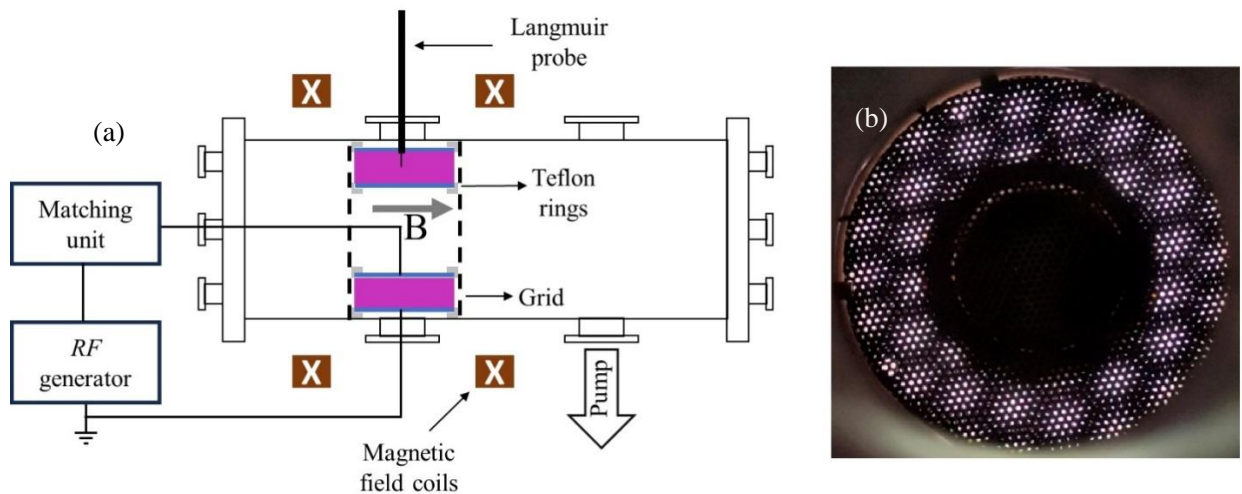

Fig. 1. (a) Schematic of the annular coaxial cylindrical CCP discharge set-up with two magnetic field coils and RF power connection scheme. (b) Photograph of the annular plasma produced in argon discharge taken from the central axial view port.

## 2.2 Magnetic field coils

Axial magnetic field is created by using two magnetic field coils arranged in Helmholtz configuration. A single core multi-strand copper cable of 16 mm$^2$ cross-section is used to fabricate two similar electromagnets having 8 layers and 8 turns. Outer frame to the coils is provided by using Hylem sheets. The inner radius of the electromagnet is kept equal to distance between them i.e., 28.5 cm. Using a Gaussmeter, the magnetic field strength is measured along the axis of the chamber by varying the current from 10 A to 50 A. As shown in Fig. 2 (a), the value of magnetic field strength increases uniformly by increasing the current applied to the electromagnets. It is observed that this arrangement of the field coils produces the region of approximately uniform magnetic field up to the distance of 28 cm between both the coils as shown in Fig. 2 (a). The region bounded by two vertically dotted lines in Fig. 2(a) shows the region of uniform magnetic field. Annular coaxial electrode assembly system is placed in this uniform axial field region inside the vacuum chamber. Fig. 2(b) shows the schematic diagram of the arrangement of electrodes in our CCP discharge source. Since the cylindrical electrodes are powered, the electric field is in the radial direction and the magnetic field due to the electromagnets is in the axial direction. In this work, the discharge gap (d) is 5 cm (0.05 m) and the voltage applied to the electrode (V) varies from ~20 V to ~70 V. So, the radial electric field varies from ~1 V/m to ~3.5 V/m. The magnetic field has increased up to 60 G. The combined effects of the electric and magnetic fields generate E×B drift in azimuthal direction as shown in Fig. 2(b). The electrons are trapped in the region bounded by the two cylindrical electrodes due to E×B drift, thereby reducing the radial loss of plasma and increasing the overall plasma density.

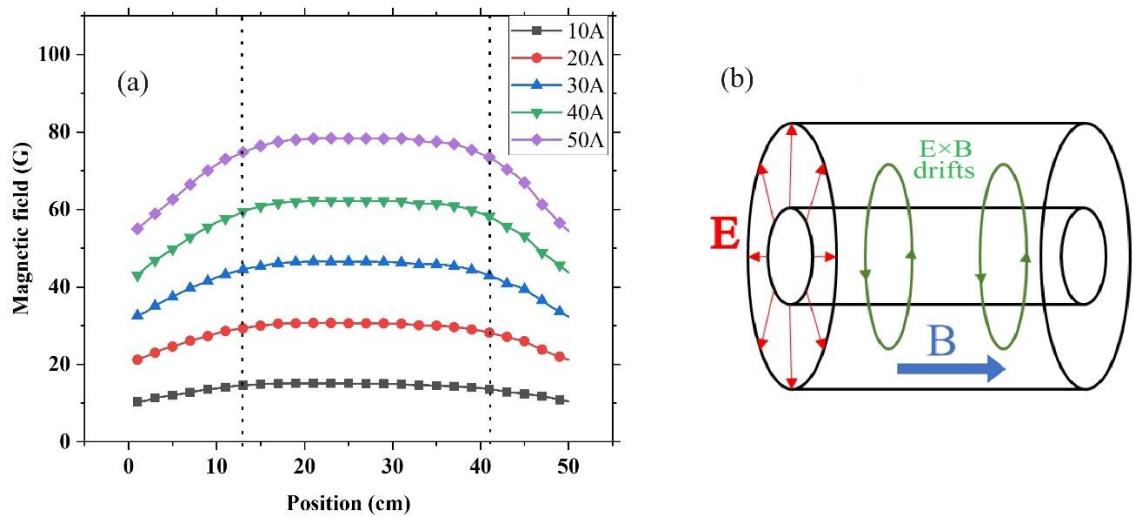

Fig. 2. (a) The variation of the magnetic field along the axis of the chamber for different values of current applied to the electromagnets. The region bounded by two vertical dotted lines shows the region of nearly uniform magnetic field. (b) Schematic diagram showing the arrangement of electrodes in which $E \times B$ forces is generated in azimuthal direction due to radial electric field and axial magnetic field.

## 2.3 Diagnostics

An RF-compensated Langmuir probe used for the measurement of plasma parameters like density, temperature and potential is designed and fabricated in-house. The Langmuir probe consists of a cylindrical tip of 0.1 mm diameter and 6 mm length and an auxiliary electrode of 1 mm diameter and 12 mm length as shown in Fig. 3(a). The presence of magnetic field requires the probe tip to be inserted perpendicular to the field direction in order to expose the maximum effective collection area to the plasma [61,62], and hence the Langmuir probe is designed with special care for the radial measurement of plasma parameters. Therefore, to optimize the RF compensation, the auxiliary electrode is kept as nearest as possible to the probe tip as the separation between probe tip and auxiliary electrode affects the RF compensation because for larger separation, the auxiliary electrode cannot sense the same RF oscillations as the probe tip. Sheath capacitance ($C_{sh}$) for both main probe tip and auxiliary electrode is deduced from equation (1) and is plotted as shown in the Fig.3 (b). [63].

$$C_{sh} = \frac{A_p \epsilon_0}{\sqrt{2} \lambda_d} \frac{(1+2\eta)^{-1/2} - e^{-\eta}}{\left[(1+2\eta)^{1/2} + e^{-\eta} - 2\right]^{1/2}} \quad (1)$$

Here, $A_p$ is contact area, $\epsilon_0$ is the permittivity of free space, $\lambda_d$ is the Debye length and $\eta = -\frac{e(V-V_p)}{k_b T_e}$, where $V$ is the voltage applied to the probe and $V_p$ is the plasma potential. The sheath capacitance of the auxiliary electrode and the probe tip are calculated by equation (1) and plotted in Fig.3 (b). The auxiliary electrode has sheath capacitance larger than ten times sheath capacitance of the probe tip. Since the sheath capacitance is directly proportional to surface area, the size of the auxiliary electrode is kept large to increase the coupling of RF voltage to the probe. Auxiliary electrode is coupled to the probe through a 0.1 nano Farad (nF) capacitor. The probe tip is connected to the series of self-resonating inductors alternately (two inductors with the inductance of 22 µH having self-resonating frequency of 13.56 MHz & additional two inductors of inductance 7 µH with self-resonating frequency 27.12 MHz) to provide high impedance at 13.56 MHz and its harmonics. These arrangements provide very high impedance ~100 kΩ and ~80 kΩ at 13.56 MHz and at 27.12 MHz respectively as shown in Fig. 3(c) and satisfies the condition $Z_f \geq \frac{Z_p e V_{rf}}{T_e}$ [64]; where $Z_f$ is probe impedance, $Z_p$ is plasma to probe impedance and $V_{rf}$ is *RF* potential of plasma to ground.

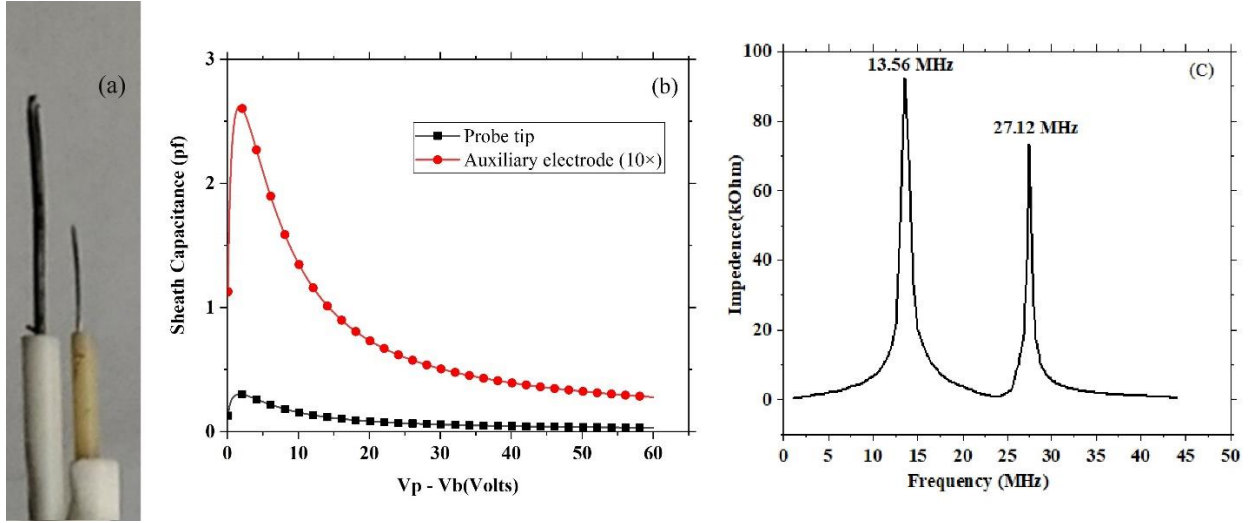

Fig. 3. Various parts of the RF compensated Langmuir Probe showing (a) the probe tip and auxiliary electrode; (b) sheath capacitance at both the probe tip and auxiliary electrode; (c) probe impedance at excitation frequency (13.56 MHz) and its second harmonics (27.12 MHz).

**2.4 Circuit for EEDF measurement by Second Harmonic Technique (SHT)**

The measurements of EEDF by numerical differentiation of I-V characteristics of the Langmuir probe data come under the noise since the noise also gets amplified upon differentiation. In order to reduce these noises, a second harmonic technique (SHT) is used for the direct measurement of the second derivative of I-V curve from the plasma without performing the actual numerical differentiation. Therefore, this method provides more appropriate results as compared to the usual EEDF measurement from the I-V characteristics. In this technique, a high frequency low amplitude sine signal is superimposed to the DC bias signal [65–67].

$$I_e(V_{bias}) = I_e(V_{dc} + V_{ac}\sin 2\pi ft) \qquad (2)$$

Where $V_{dc}$ and $V_{ac}$ are the amplitudes of DC bias and high frequency sine signal. By applying Tailor expansion to equation (2) and by separating the harmonic components, we get

$$\begin{aligned}I_e &= I_e(V_{dc}) + \frac{V_{ac}^2}{4}\frac{d^2 I_e}{dV^2} + \frac{V_{ac}^4}{64}\frac{d^4 I_e}{dV^4} + \left[V_{ac}\frac{dI_e}{dV} + \frac{V_{ac}^3}{8}\frac{d^3 I_e}{dV^3} + \cdots\right]\sin 2\pi ft \\ &\quad - \left[\frac{V_{ac}^2}{4}\frac{d^2 I_e}{dV^2} + \frac{V_{ac}^4}{48}\frac{d^4 I_e}{dV^4} + \cdots\right]\cos 4\pi ft + \cdots\end{aligned} \qquad (3)$$

From equation (3), we can see that the amplitude of second harmonics is proportional to the second derivatives of I-V characteristics. Although the second harmonic component contains second order as well as the higher order derivatives, the amplitude of ac signal is kept low to neglect higher order derivative terms. The second derivative of I-V characteristics obtained from the second harmonic technique is substituted in Druyvesteyn relation given in equation (4) to determine the EEDF.

$$f_e(E) = \frac{4}{e^2 A}\left(\frac{m}{2}\right)^{\frac{1}{2}} E^{\frac{1}{2}} \frac{d^2 I_e}{dV^2} \qquad (4)$$

Here, $E = eV$, where $V = V_p - V_b$. $V_p$ is plasma potential and $V_b$ is biasing potential and $A$ is the area of the probe tip. Plasma density and electron temperature can be obtained from EEDF using equation (5) and (6)

$$n_e = \int_0^\infty f_e(E) dE \qquad (5)$$

$$T_{eff} = \frac{2}{3 n_e} \int_0^\infty E f_e(E) dE \qquad (6)$$

An electronic circuit consisting of a summing amplifier for superimposing a low amplitude high frequency sine signal to the high amplitude low frequency ramp signal by using differential isolation amplifier for the differential measurement (current measurement) across a known resistor is designed and fabricated in house. In this circuit, high voltage amplifier IC (PA85) is employed in unity gain summing mode. It provides a very stable summing signal of high voltage ramp supplied by the high voltage amplifier and the sine signal obtained from the lock-in-amplifier to the Langmuir probe through current measurement resistor. A differential isolation amplifier (AD215) is used for the measurement of the current drawn by the Langmuir probe by measuring the voltage drop across the resistor in unity gain differential mode. AD215 has high working bandwidth of 120 kHz and provides a very high common mode rejection ratio (CMRR ~ 100dB). The differential signal obtained is supplied to the lock-in-amplifier that locks the second harmonic frequency of the sine signal because the amplitude of second harmonic signal is equivalent to the second derivative of the probe current. The arrangement of the various components of circuit used for the measurement of I-V characteristics and its second harmonics in this work by using Langmuir probe is shown in Fig. 4.

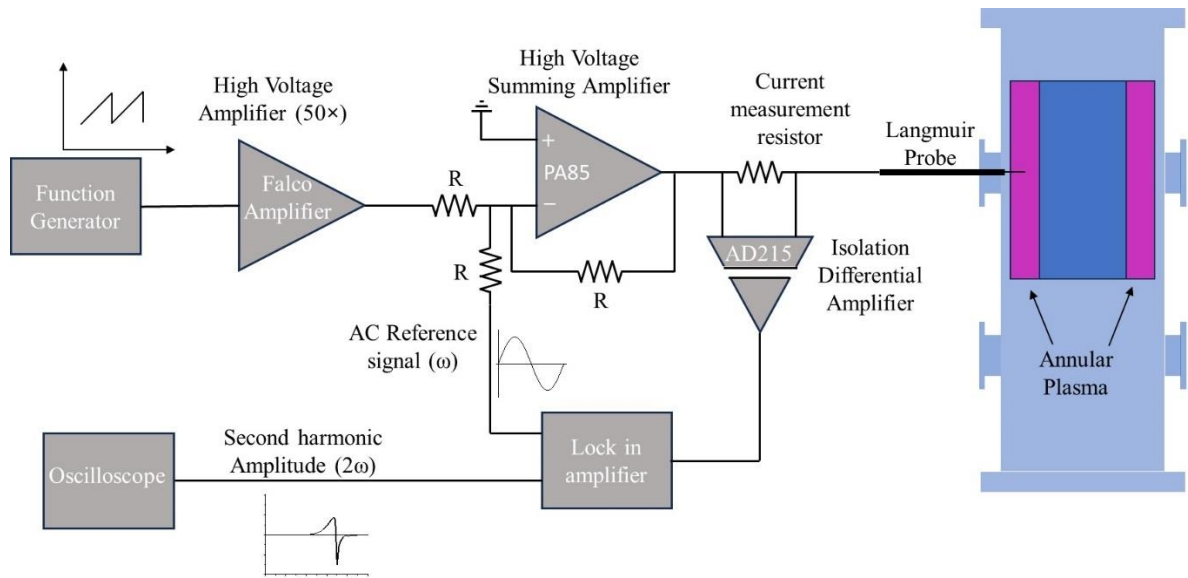

Fig. 4. Block diagram of the circuit consisting of the various electronic components used for the measurement of I-V characteristics of the Langmuir probe and its double derivative using second harmonic technique (SHT).

## 3. Results and discussions

In this experiment, the vacuum chamber is evacuated up to a base pressure of $5\times10^{-6}$ mbar by using diffusion pump backed by a two-stage rotary vane pump. High purity argon gas is injected into the chamber with the help of a Pfeiffer made gas dosing valve so as to obtain a working pressure of 1 Pa. The plasma is produced in the annular region bounded between two cylindrical electrodes. The inner cylindrical electrode is powered by using 13.56 MHz RF generator and matching network, and the outer cylindrical electrode is grounded. The characterization of plasma is performed by using an RF-compensated Langmuir probe by varying the RF power and magnetic field strength. The EEDF measurements are performed at 1 Pa and 3 Pa gas pressure. The probe is inserted radially into the chamber and fixed at the centre of the discharge (as shown in Fig.1) in order to minimize the effect of magnetic field on the measurement and also to expose maximum collection area of the tip to the plasma. EEDFs are directly obtained for different cases by using second harmonic technique (SHT) as discussed in section 2.4. In the experiments, the ramp voltage signal is set from -80 V to 120 V with a frequency of 10 Hz, and the amplitude and frequency of superimposed AC ($V_{ac}$) signal are 2 V and 2 kHz respectively. Electron density and temperature are obtained directly from EEDF measurement by using equations (5) and (6).

Fig. 5 (a) shows the variation of plasma density and electron temperature versus applied RF power from 20-100 Watts at a constant working pressure of 1 Pa. The error bars on the plot are estimated based on multiple experimental measurements. The measured plasma density at 20 Watt is ~4.8 × 10$^{15}$ m$^{-3}$ and the density shows approximately 100% rise when RF power is increased to 100 Watt i.e., reaches to ~8.50 × 10$^{15}$ m$^{-3}$. An increase in the plasma density with RF power is attributed to a rise in the sheath voltage. Thus, the electrons absorb more power from the RF field through stochastic heating mechanism leads to higher plasma density by direct and multiple step ionization processes with the background neutrals. In Fig. 5 (a), we have observed that the measured electron temperature is ~3 eV and remains approximately unchanged with RF power. In low-temperature plasmas, electron temperature is dependent on the power balance involving electron creation and losses mechanism. Hence, the electron temperature is dependent on the product of gas pressure (p) and characteristic size (L) of the plasma. Thus, the electron temperature does not vary significantly with the applied RF power [68]. Fig. 5 (b) shows the variation of plasma potential estimated from IV curve when the RF power is increased from 20 Watt to 100 Watt. An increase in the plasma potential is observed from 53 V at 20 Watt to 110 V (~107%) when the RF power is increased to 100 Watt. This is similar to electron density rise with RF power i.e., both of these parameters increase by two-fold with 5 times increase in RF power. An increase in the plasma potential with RF power is to confine the high energy electrons produced in the discharge.

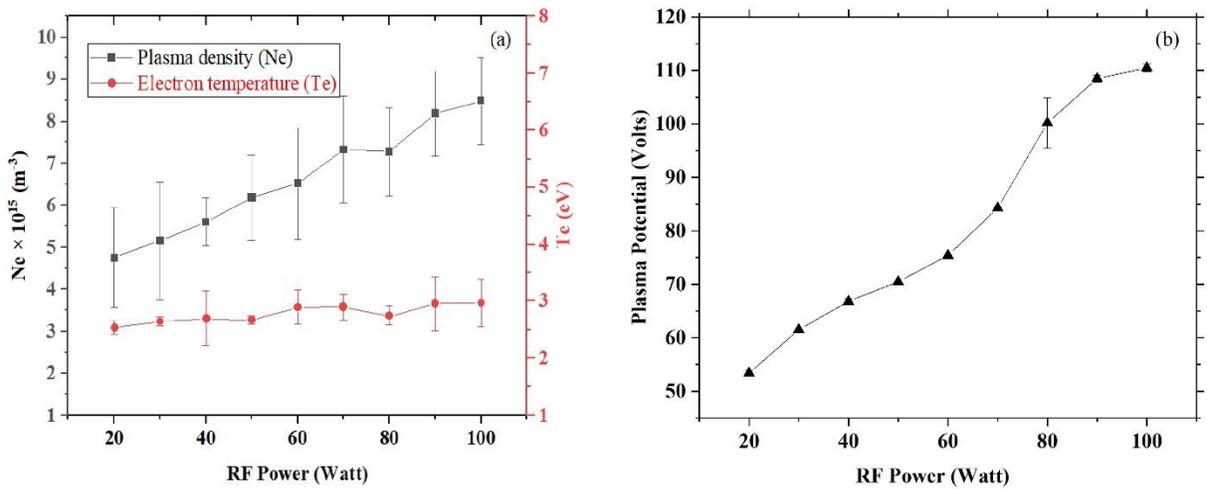

Fig. 5. Plot of the variation of (a) plasma density (black) & electron temperature (red) and (b) plasma potential with RF power when varied from 20-100 Watt at the constant working pressure of 1 Pa.

The experimental results are verified by comparing the measured values of electron density and temperature with the values computed by using global model based on particle and power balance equations [8]. The plasma density and electron temperature can be calculated by this model by incorporating the system parameters and operational conditions like discharge dimensions, working pressure, RF power and background neutral gas density. However, the plasma parameters calculated by using particle and power balance equations only provide the spatially averaged results, independent of the type and nature of the discharge. In fact, these equations only give the trends or the variation of the plasma parameters with the operating condition rather than the exact values. This model has been previously used by many researchers to estimate the approximate values of plasma parameters like density and temperature [8,69–71].

The particle balance equation for the calculation of electron temperature in the discharge is given by:

$$\frac{k_{iz}}{u_B} = \frac{1}{n_g d_{eff}}. \quad (7)$$

Here, $k_{iz}$ is ionization coefficient for argon atom, $u_B$ is the Bohm velocity given by $u_B = \sqrt{\frac{KT_e}{m_i}}$, where $m_i$ is the mass of positive Ar$^+$ ions, $n_g$ is neutral gas density and $d_{eff}$ is the effective plasma size. The value of $k_{iz}$ is obtained from the Lieberman's "Principals of Plasma Discharges and Materials Processing" [8]. Electron density is calculated by using power balance equation, given by:

$$n_0 = \frac{P_{abs}}{e u_B A_{eff} E_T} \quad (8)$$

In equation (8), $P_{abs}$ is the discharge power; $A_{eff}$ is the effective area for electron loss. $E_T$ is the total energy consumed from creation upto the loss of an electron-ion pair. $E_T$ is defined as the sum of mean kinetic energy lost per electron lost $E_e = 2T_e$, mean kinetic energy lost per ion lost i.e $E_i = V_s + T_e/2$. Here, $V_s = \frac{T_e}{2} \ln\left(\frac{m_i}{2\pi m_e}\right)$ is sheath potential and collisional energy lost for generation of an electron-ion pair ($E_c$) i.e $E_T = E_e + E_i + E_c$.

The expression for the collisional energy ($E_c$) as the function of electron temperature is given by [8,71,72].

$$E_c = E_{iz} + \sum_i E_{ex,i} \frac{k_{ex,i}}{k_{iz}} + \frac{k_{el}}{k_{iz}} \frac{3m_e}{m_i} T_e \quad (9)$$

Here, $E_{iz}$ is the ionization energy of argon, $E_{ex,i}$ is the minimum energy (threshold energy) required for $i^{th}$ excitation of argon atom, $k_{ex,i}$ and $k_{el}$ are the rate coefficients for the $i^{th}$ excited state and elastic collision respectively. All the rate coefficients and energies used for the calculation of $Ec$ as the function of $Te$ are listed in the Table 1 and Table 2 [8,73,74]. Using equation (9), $Ec$ is calculated for different values of $Te$ as and is plotted in as shown in Fig.6.

Table. 1: Excitation rate coefficients of Argon

| Sr. No. | Final state | Rate coefficient (m³s⁻¹) | Threshold E (eV) | Ref. |
|---|---|---|---|---|
| 1. | $^1P_1$ | $k_1 = 2.72 \times 10^{-16} \exp(-12.14/T_e)$ | 11.8 | [73] |
| 2. | $^3P_0$ | $k_2 = 1.35 \times 10^{-15} \exp(-12.42/T_e)$ | 11.7 | [73] |
| 3. | $^3P_1$ | $k_3 = 1.91 \times 10^{-15} \exp(-12.60/T_e)$ | 11.6 | [73] |

| 4. | $^3P_2$ | $k_4 = 5.02 \times 10^{-15} \exp(-12.64/T_e)$ | 11.5 | [73] |
| 5. | 4p | $k_5 = 2.12 \times 10^{-14} \exp(-13.13/T_e)$ | 13.2 | [74] |
| 6. | 4s, 4s' | $k_6 = 1.45 \times 10^{-14} \exp(-12.96/T_e)$ | 11.8 | [74] |
| 7. | $5s, 3\bar{d}, 5s', 3d'$ | $k_7 = 1.22 \times 10^{-14} \exp(-12.96/T_e)$ | 14.2 | [74] |
| 8. | $4d, 6s, 4\bar{d}, 4d', 6s', 5d, 7s, 5\bar{d}$ | $k_8 = 7.98 \times 10^{-15} \exp(-19.05/T_e)$ | 15.0 | [74] |
| 9. | Higher state | $k_9 = 8.29 \times 10^{-15} \exp(-18.14/T_e)$ | 15.5 | [74] |

Table. 2: Ionization and elastic scattering rate coefficients

| Sr. No. | Final state | Rate coefficient | Threshold E (eV) | Ref. |
|---|---|---|---|---|
| Ionization | e + Ar → Ar$^+$ + 2e | $k_{iz} = 2.34 \times 10^{-14} T_e^{0.59} \exp(-17.44/T_e)$ | 15.76 | |
| Elastic | e + Ar → Ar + e | $\ln(k_{el}) = -31.3879 + 1.609 \ln(T_e)$ $+ 0.0618 (\ln(T_e))^2$ $+ 0.1171 (\ln(T_e))^3$ | | [8] |

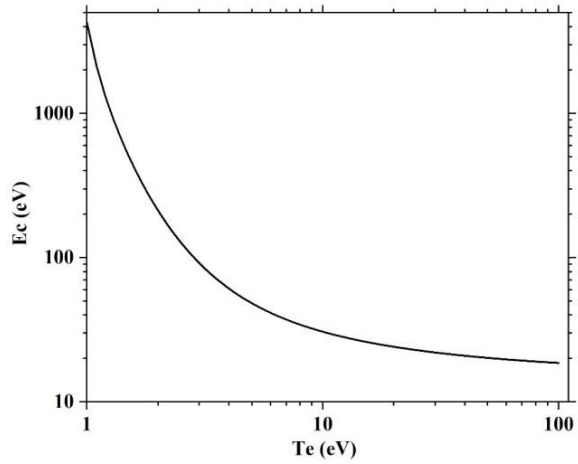

Fig.6. Plot of the variation of collisional energy lost per electron-ion pair created (*Ec*) as the function of electron temperature (*Te*) for Argon discharge.

Equations (7) and (8) are used to calculate the plasma density electron temperature for different values of applied RF power. The calculated values of electron density and temperature are compared with the experimentally measured values as shown in Fig.7. It is observed that the density obtained from equation (8) is larger than the density measured by using RF compensated Langmuir probe. This is due to the fact that, the powers recorded in the experiments are the power displayed on the screen of RF generator. However, the actual power delivered to the plasma by the electrode is much less than the displayed power due to losses in the matching unit, cables, connectors, electrode, etc. In this work, the power transfer efficiency is approximately 50 percent. However, in the global model, no such losses are taken into consideration. Therefore, it is observed that the experimentally measured plasma density is less than the density calculated by using the global model. However, both in the global model and in the experiment; plasma density gradually increases with the applied RF power. Electron temperature calculated by using equation (7) comes out to be approximately 3 eV as shown in Fig.7 (b). The calculated value of electron temperature is constant with RF power and shows good agreement with the experimental results.

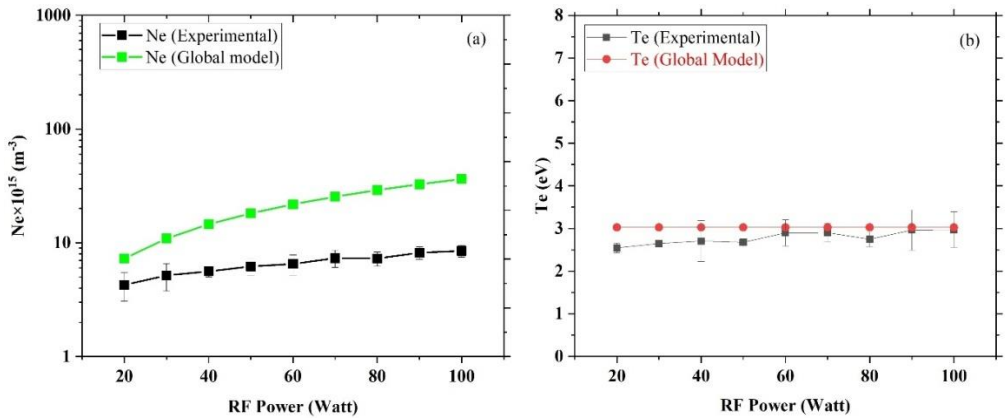

Fig. 7. Plot showing the comparison of the variation of experimental and calculated values of (a) plasma density (*Ne*) and (b) electron temperature (*Te*) with RF power when varied from 20 Watt to 100 Watt at the working pressure of 1 Pa.

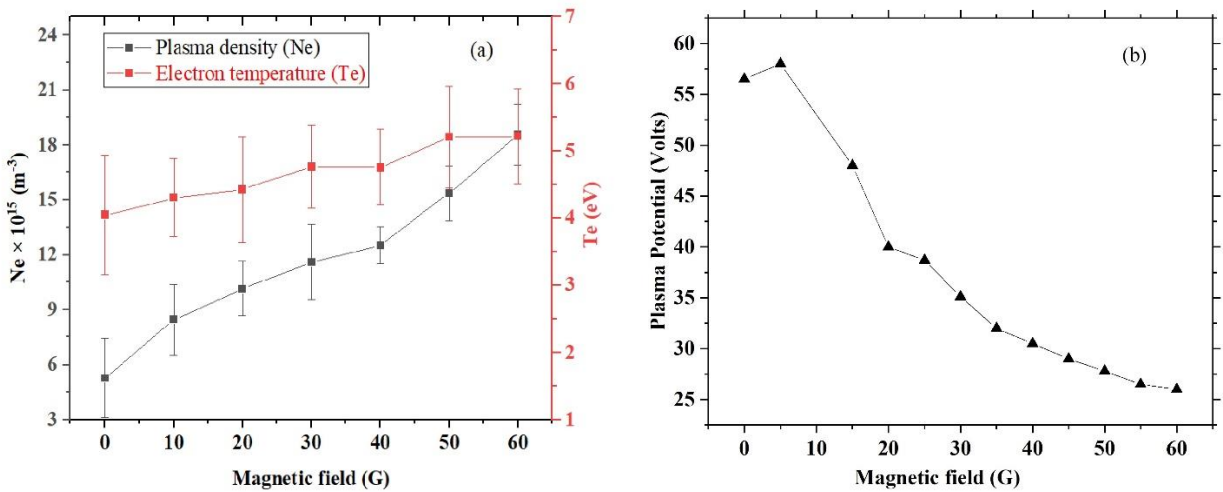

Fig. 8. Plot of the variation of (a) plasma density (black) & electron temperature (Red), and (b) plasma potential with magnetic field at the applied RF power of 20 Watt when the working pressure is 1 Pa.

Fig. 8 (a) shows the variation of electron density and electron temperature as a function of magnetic field for a constant RF power of 20 Watt and working pressure of 1 Pa. The measured plasma density increases linearly up to $1.8 \times 10^{16}$ m$^{-3}$ when the magnetic field strength gradually increased to 60 G i.e., a three-fold rise is observed when compared to non-magnetized condition. The ionization percentage varies from ~0.5 percent to ~1.8 percent and the electron neutral collision frequency is ~ 5 MHz. As the magnetic field is applied, the hot electrons are confined due to azimuthal force ($v_r \times B_z$), which drive electrons to follow gyro-motion around the magnetic field lines thereby reducing the electron loss to the powered and grounded electrodes. For a magnetic field strength of 10 G, electrons with temperature of 15 eV the calculated electron gyro-frequency ($f_{ce}$) and gyro-radius ($r_{ce}$) are 28 MHz and 13 mm respectively. Similarly, for an argon ion Ar$^+$ with temperature of 0.05 eV, gyro-frequency ($f_{ci}$) and gyro radius ($r_{ci}$) are 0.38 kHz and 200 mm. Hence, the hot electrons with energies required to ionize the neutral particles are magnetized and thus enhancing the plasma density by electrons colliding with the background neutral atoms. On the other hand, ions remain unmagnetized as they have gyro-radius larger than the discharge gap (50 mm). Therefore, with an increase in the magnetic field strength, the high energy electron population with energies above ionization potential increase, thereby increasing overall plasma density linearly with magnetic field. The electron temperature also slightly increases with magnetic field as shown in Fig. 8 (a). When electrons get magnetized, their gyrofrequency becomes slightly higher than the RF excitation frequency. Therefore, the electrons follow both the electric field oscillations due to applied RF power as well as gyro motion due to applied magnetic field; these enhance the impact of Ohmic heating resulting in the slight increment of electron temperature. Fig .8 (b) shows the variation of plasma potential as the function of magnetic field when varied from 0 to 60 G. The plasma potential drops rapidly from 56 V at 0 G to 26 V (approximately 53%) at 60 G. This is due to the fact that bulk plasma bulk attains higher potential to reduce the electron loss and to keep the plasma quasi-neutral but the magnetic field reduces the loss of electrons to the grounded electrode thereby confining the electrons inside the plasma bulk. This effect reduces the ambipolar bulk plasma potential with respect to ground. Similar results have been reported in Singh *et al*. [75].      In the subsequent section, we focus on the study of EEDF for different operating parameters like RF power, gas pressure and axial magnetic field strength. Fig. 9 (a) shows the variation of EEDF with RF power at a constant gas pressure of 1 Pa in the absence of external magnetic field (B = 0 G), Fig. 9 (b) shows the effect of gas pressure on the EEDF at a fixed RF power of 100 Watt and B = 0 G, and Fig.9 (c) shows the variation of EEDF for different magnetic field strengths at a constant gas pressure and RF power of 1 Pa and 20 Watt respectively.

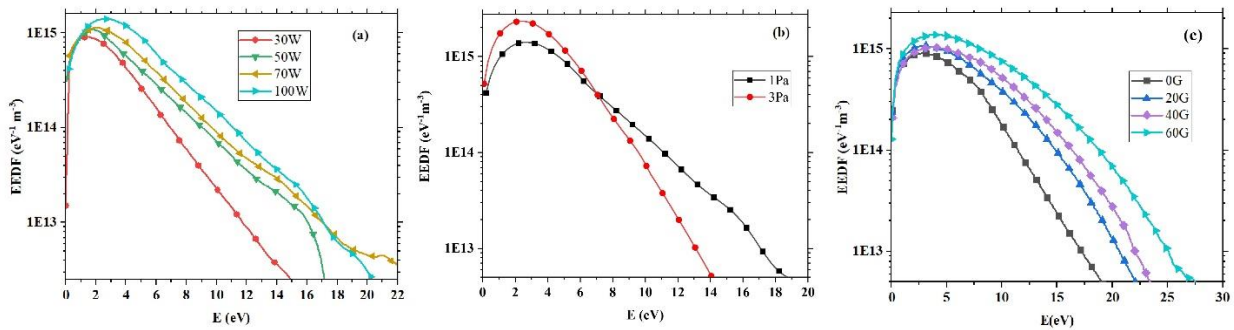

Fig. 9. Plot showing the variation of EEDF with (a) RF power at a constant gas pressure of 1 Pa and B = 0 G, (b) pressure at the fixed RF power of 20 Watt and B = 0 G, and (c) varying magnetic field strength up to 60 G for constant pressure and RF power of 1 Pa and 20 Watt respectively.

From Fig. 9 (a), which shows that the variation of EEDF with RF power, it can be seen that the area under the EEDF curve increases proportionately when the externally applied RF power is increased. However, an increase in the applied RF power does not influence much on the shape of EEDF. A nearly Maxwellian EEDF is observed for all the applied RF powers. The increase of RF power enhances the overall density of the plasma, therefore the overall area under EEDF curve increases. As discussed above, the applied RF power has a slight influence on values of electron temperature, therefore when the applied RF power is increased, the density of the electrons with different energies increases by equal proportion and hence the shape of EEDF curve remains approximately unchanged as shown in Fig 9 (a). Fig. 9 (b) shows the effect of gas pressure on the shape of EEDF at 100-Watt RF power when there is no axial magnetic field is applied. When the pressure increases from 1 Pa to 3 Pa, the background neutral gas density increases that leads to more frequent electron-neutral collisions due to a reduction in the electron neutral mean free path. Thus, higher collision rate results in the depletion of high energy electrons because they are more likely to lose their energy through collision and increase the density of low energy electrons. More collisions enhance the ionization and hence the overall plasma density increases with pressure. Fig. 9 (c) shows the variation of EEDF for different values of magnetic field at a constant gas pressure (1 Pa) and RF power (20 W). Increasing magnetic field strength from 0 G to 60 G, the confinement of electrons in the bulk plasma increases which enhances the collision and hence increases the overall plasma density. Due to this increase of plasma density with the increase of axial magnetic field, the area under the EEDF curve increases. In the presence of magnetic field, the electrons follow gyro-motion around the field lines. But the hot electrons have larger gyro-radius and hence they are more likely to lose towards the walls of the chamber or on to the electrodes when compared to cold electrons. Additionally, the high energy electrons losses their energy by ionization and in this process, the density of the low and mid energy range electrons will increase. Thus, with an increase in the magnetic field strength, an increase in the density of hot electrons is less as compared to that of the mid and cold electrons. Therefore, the gradual transition of EEDF from Maxwellian type to Druyvesteyn type occurs when the axial magnetic field is increased.

4. **Conclusion**

A systematic experimental study is performed in a novel symmetric magnetized cylindrical CCP discharge using an RF-compensated LP. The plasma source design includes two co-axial cylinders in which the inner cylinder is powered, the outer cylinder is grounded, and the plasma is produced in the region bounded by these two cylinders. Uniform axial magnetic field strength up to 80 G is applied using a set of electromagnet coils. The radial electric field along with the axially uniform magnetic field provide plasma confinement in the azimuthal direction due to E×B drift. An RF-compensated Langmuir probe is constructed and is used for the measurement of localized plasma parameters including EEDF in the plasma bulk. A data acquisition system along with the associated electronics is designed for the direct measurement of EEDF based on the second harmonic technique. It is observed that the plasma density increases with RF power i.e., when the power is increased from 20 Watt to 100 Watt, the plasma density increases approximately up to 100 %, however the electron temperature remains approximately unchanged with the RF power. It is also observed that the application of axial magnetic field increases the plasma density. When the magnetic field increases from 0 G to 60 G, the plasma density increases by three-fold. The electron temperature also shows a slight increase with the magnetic field strength. The results show that the magnetic field reduces the radial plasma losses thereby increasing the overall density. To validate our result, the density and temperature measured by using LP is compared with the values computed by global model involving particle and power balance. Comparison shows a good correlation between experimental results and computed values. The measured and calculated electron temperature is approximately the same whereas the measured density is found to be lower than the calculated density. The reason for lower values of experimental plasma density as compared to the calculated values is due to the loss of RF power in the matching network, cables, electrode, connectors, etc in the experiment. EEDFs are measured in the system by varying the RF power, magnetic field and gas pressure. It is observed that the shape of EEDF remains approximately unchanged with the RF power because of the proportionate increase of low, mid and high energy electrons with power. However, a transition in the shape of EEDF from Maxwellian type to Druyvesteyn type is observed when the magnetic field is applied. Depletion of high energy electrons and the increase of low energy electrons are observed when the gas pressure is increased because high energy electrons lose their energy through collision thereby increasing the overall plasma density. This annular plasma source is developed with the future plans to study the effect of driving frequencies (2 MHz, 13.56 MHz, 27.12 MHz and 60 MHz) on the power coupling and electron heating mechanism in the symmetric CCP discharge. A comparison between PIC simulation and experiments will be subject of future publication.

**Acknowledgements**

The authors acknowledge Ravi Ranjan Kumar for his valuable technical support during setting up the experiments.

This work was supported by the Department of Atomic Energy, Government of India, and Science and Engineering Research Board (SERB), Core Research Grant No. CRG/2021/003536.